\documentclass[prb,aps,twocolumn,showpacs,superscriptaddress,preprintnumbers,floatfix,amsmath,amssymb]{revtex4-1}

\usepackage{amssymb}
\usepackage{graphicx}
\usepackage{dcolumn}
\usepackage{bm}
\usepackage{color}
\hfuzz=\maxdimen
\begin{document}

\title{Giant anisotropy in superconducting single crystals of CsCa$_2$Fe$_4$As$_4$F$_2$}

\author{Zhi-Cheng Wang}
\affiliation{Department of Physics, Zhejiang University, Hangzhou 310027, China}

\author{Yi Liu}
\affiliation{Department of Physics, Zhejiang University, Hangzhou 310027, China}

\author{Si-Qi Wu}
\affiliation{Department of Physics, Zhejiang University, Hangzhou 310027, China}

\author{Ye-Ting Shao}
\affiliation{Department of Physics, Zhejiang University, Hangzhou 310027, China}

\author{Zhi Ren}
\affiliation{Institute of Natural Sciences, Westlake Institute for Advanced Study, Westlake University, Hangzhou 310064, China}

\author{Guang-Han Cao} \email[Correspondence should be sent to: ]{ghcao@zju.edu.cn}
\affiliation{Department of Physics, Zhejiang University, Hangzhou 310027, China}
\affiliation{State Key Lab of Silicon Materials, Zhejiang University, Hangzhou 310027, China}
\affiliation{Collaborative Innovation Centre of Advanced Microstructures, Nanjing University, Nanjing 210093, China}

\date{\today}

\begin{abstract}
CsCa$_2$Fe$_4$As$_4$F$_2$ is a newly discovered iron-based superconductor with $T_\mathrm{c}\sim$ 30 K containing double Fe$_2$As$_2$ layers that are separated by insulating Ca$_2$F$_2$ spacer layers. Here we report the transport and magnetization measurements on CsCa$_2$Fe$_4$As$_4$F$_2$ single crystals grown for the first time using the self flux of CsAs. We observed a huge resistivity anisotropy $\rho_c(T)/\rho_{ab}(T)$, which increases with decreasing temperature, from 750 at 300 K to 3150 at 32 K. The $\rho_c(T)$ data exhibit a non-metallic behavior above $\sim$140 K, suggesting an incoherent electronic state at high temperatures due to the dimension crossover. The superconducting onset transition temperature in $\rho_{ab}$ is 0.7 K higher than that in $\rho_c$, suggesting two-dimensional (2D) superconducting fluctuations. The lower and upper critical fields also show an exceptional anisotropy among iron-based superconductors. The $H_{c1}^\bot(T)$ data are well fitted using the model with two $s$-wave-like superconducting gaps, $\Delta_1(0)=6.75$ meV and $\Delta_2(0)=2.32$ meV. The inter-plane coherence length $\xi_c(0)$ is $3.6$ \AA, remarkably smaller than the distance between conducting layers (8.6 \AA), consolidating the 2D nature in the title material.
\end{abstract}


\maketitle

\section{INTRODUCTION}

Fe-based high-temperature superconductors (FeSCs) are structurally characterized by the anti-fluorite-type Fe$_2$$X_2$ ($X=$ As, Se) layers~\cite{johnston,jh,hosono-pc} which serve as the superconductingly active motif. Most FeSCs discovered consist of either separate Fe$_2$$X_2$ monolayers represented by the ``1111"-type compounds or, infinite Fe$_2$$X_2$ layers exemplified by the ``11"- and ``122"-type materials~\cite{jh}. Recently, we discovered a series of double-Fe$_2$As$_2$-layer FeSCs, $Ak$Ca$_2$Fe$_4$As$_4$F$_2$ ($Ak$ = K, Rb, Cs)~\cite{KCa2Fe4As4F2,ACa2Fe4As4F2} and $AkLn_2$Fe$_4$As$_4$O$_2$ ($Ak$ = K, Rb, Cs; $Ln$ = Nd, Sm, Gd, Tb, Dy, Ho)~\cite{RbGd2Fe4As4O2,RbLn2Fe4As4O2,ALn2Fe4As4O2}, which resemble the cuprate superconductors with double CuO$_2$ sheets. Those so-called 12442-type FeSCs are actually resulted from the intergrowth of 1111- and 122-type iron arsenides. Figure \ref{XRD}(a) shows the crystal structure of one of the family members, CsCa$_2$Fe$_4$As$_4$F$_2$, which is also the research object of the present study. As is seen, the Cs$^+$-cations connected double Fe$_2$As$_2$ layers are separated by the insulating Ca$_2$F$_2$ block. Note that the alkali-metal-containing 122 block is nominally hole doped with 0.5 holes/Fe, while the 1111 block is nominally undoped. Consequently, the 12442-type compounds are all hole doped by themselves at a level of 0.25 holes/Fe, which makes them superconducting at $T_c=$ 28$-$37 K without extrinsic doping~\cite{KCa2Fe4As4F2,ACa2Fe4As4F2,RbGd2Fe4As4O2,RbLn2Fe4As4O2,ALn2Fe4As4O2}. With an extrinsic electron doping through Co/Fe substitution, superconductivity gradually disappears, accompanied with a sign change in Hall coefficient~\cite{12442.Hosono}.

Our previous magnetoresistance measurements on polycrystalline samples of 12442-type superconductors show that the superconducting
transition is severely broadened with pronounced
tails under external magnetic fields, similar to the case in quasi-two-dimensional (quasi-2D) cuprates~\cite{Blatter1994}. Consequently, there exists a large gap between the upper critical field $H_{c2}(T)$ and the irreversible field $H_\mathrm{irr}(T)$~\cite{KCa2Fe4As4F2,ACa2Fe4As4F2,RbGd2Fe4As4O2}, suggesting a pronounced vortex-liquid phase because of the enhanced two dimensionality as also indicated by the first-principles studies~\cite{12442.Hosono,12442LDA1.WangGT,12442LDA2.WangGT,12442LDA3}. Furthermore, the initial slope of $\mu_0H_{c2}(T)$ achieves $-$16.5 T/K for the RbGd$_2$Fe$_4$As$_4$O$_2$ polycrystals~\cite{RbGd2Fe4As4O2}, which implies small superconducting coherence lengths. So far, all the related works~\cite{KCa2Fe4As4F2,ACa2Fe4As4F2,RbGd2Fe4As4O2,RbLn2Fe4As4O2,ALn2Fe4As4O2,12442.Hosono,KCa12442-muSR,CsCa12442-muSR,12442-Mossbauer} were performed on polycrystalline samples. To reveal the intrinsic anisotropic properties, nevertheless, one needs the measurements on single crystals of the 12442-type FeSCs. In this article, we report the growth and the anisotropic properties of CsCa$_2$Fe$_4$As$_4$F$_2$ single crystals. The main result shows an unexpectedly large anisotropy both in the normal state and the superconducting state, different with the common knowledge of FeSCs with relatively small anisotropy~\cite{GinzburgNumber_FeSC,Jc-Hosono_review,review.YHQ}. The strongly anisotropic behavior is reminiscence of those of most cuprate superconductors, implying the crucial role of two dimensionality for the superconductivity.

\section{EXPERIMENTAL METHODS}
Single crystals of CsCa$_2$Fe$_4$As$_4$F$_2$ were grown in the CsAs flux with a molar ratio of CsCa$_2$Fe$_4$As$_4$F$_2$ : CsAs = 1 : 14. The source materials were Cs ingot (99.5\%), Ca shot (99.5\%), Fe powders (99.99\%), As pieces (99.999\%) and CaF$_2$ powders (99\%). First, CsAs was prepared by reacting Cs and As in a sealed quartz tube at 200 $^\circ$C for 10 hours. Intermediate products of CaAs and Fe$_2$As were similarly synthesized at 750 $^\circ$C for 12 h in evacuated quartz tubes, respectively. Then the precursors were mixed with CaF$_2$ at a molar ratio of CsAs : CaAs : Fe$_2$As : CaF$_2$ = 15 : 1 : 2 : 1. The mixtures were loaded in an alumina tube, subsequently sealed in a Ta tube. The Ta tube was jacketed with an evacuated quartz ampoule. The sample-charged assembly was slowly heated to 1150 $^\circ$C, holding for 20 h. The crystals were expected to nucleate and grow up during the slow cooling down to 800 $^\circ$C at a rate of 3.75 $^\circ$C/h. Finally, shiny plate-like crystals [see Fig.~\ref{XRD}(b)] were harvested after washing away the CsAs flux with deionized water.

The single crystals was characterized by x-ray diffraction (XRD) using a PANalytical X-ray diffractometer with the CuK$_{\alpha1}$ radiation. The anisotropic electrical resistivity measurements were carried out on a Quantum Design Physical Property Measurement System (PPMS-9) adopting the conventional four-terminal method with an excitation current of 1 mA. The direct-current (dc) magnetization was measured on a Quantum Design Magnetic Property Measurement System (MPMS3) with a crystal whose dimension was $1.2\times1.8\times0.030$ mm$^3$ (0.34 mg). Note that the samples used for the above measurements were cleaved from the same piece of single crystal.

\section{RESULTS AND DISCUSSION}

\subsection{X-ray diffraction}

\begin{figure}
\includegraphics[width=8cm]{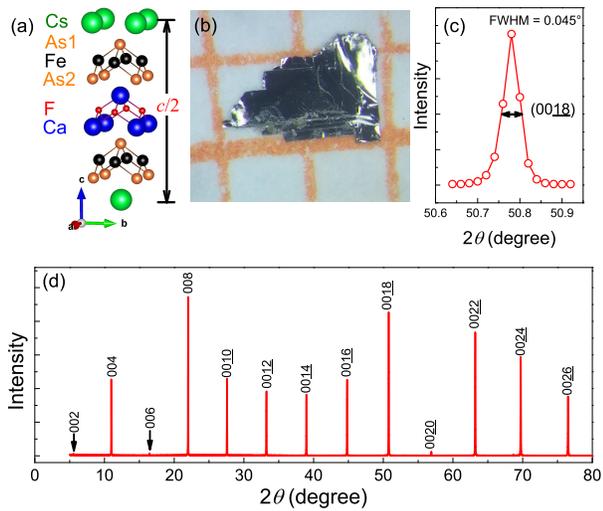}
\caption{\label{XRD}(a) Crystal structure (only half unit cell is shown) of CsCa$_2$Fe$_4$As$_4$F$_2$ which highlights the insulating Ca$_2$F$_2$ layer that separates the superconducting double Fe$_2$As$_2$ layers. (b) Photograph of a piece of single crystal on millimeter grid paper. (d) X-ray diffraction pattern ($\theta-2\theta$ scan) of the CsCa$_2$Fe$_4$As$_4$F$_2$ single crystal showing (00$l$) reflections with $l=$ even numbers exclusively. Panel (c) plots a close-up for the (00\underline{18}) reflection, from which the Full width at half maximum (FWHM) can be seen.}
\end{figure}

Figure~\ref{XRD}(d) shows a typical XRD pattern of the CsCa$_2$Fe$_4$As$_4$F$_2$ single crystal. Only (00$l$) reflections with $l$ = even numbers appear, consistent with the body centered lattice~\cite{ACa2Fe4As4F2}. One of the diffraction peaks, the (00\underline{18}) reflection, is scaled up in Fig.~\ref{XRD}(c). One sees that the peak is very sharp with a full width at half maximum (FWHM) as small as $0.045^{\circ}$, suggesting high quality of the crystal.

With the measured $d$ spacings corresponding to (00$l$) diffractions, the $c$ axis is calculated to be 32.331(3) \AA~ by a least squares fit, which is very close, yet slightly smaller than, the value of 32.363(1) \AA\ obtained from the powder XRD pattern of the polycrystalline sample~\cite{ACa2Fe4As4F2}. Here we give a plausible explanation. The polycrystalline sample could have slight Cs vacancies due to loss of Cs during the high-temperature sintering. The Cs deficiency corresponds to a higher formal valence of Fe, which would give rise to a higher $c/a$ ratio~\cite{Eu1144.Mossbauer}. As for the single crystals, which were grown in a Cs-rich flux, the Cs vacancies could be minimized. Hence the $c$ value is lower. If this is the case, the hole-doping level in the single crystals is closer to 0.25 holes/Fe, while the previous polycrystalline samples may be overdoped. Indeed, the $T_\mathrm{c}$ value of the single crystals is about 2 K higher (see below) than that of the polycrystalline samples.

\subsection{Resistivity}

\begin{figure}
\includegraphics[width=8.5cm]{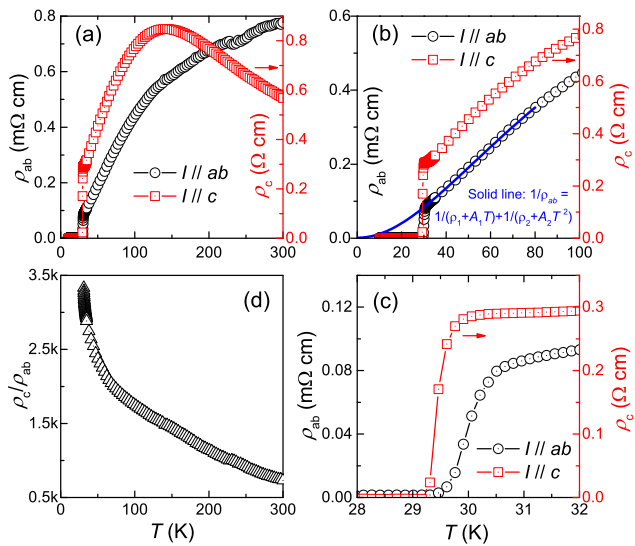}
\caption{\label{RT} (a) Temperature dependence of the in-plane ($\rho_{ab}$, with electric current flowing in the basel plane, left axis) and out-of-plane ($\rho_c$, with electric current along the $c$ axis, right axis) resistivity of CsCa$_2$Fe$_4$As$_4$F$_2$ single crystals at zero magnetic field. Panels (b) and (c) show a close-up in different temperature ranges, respectively. The solid line in (b) is the fitted result using the formula shown (see the text also). Panel (d) plots the temperature dependence of the anisotropic resistivity ratio $\rho_c/\rho_{ab}$. }
\end{figure}

\begin{figure*}
\includegraphics[width=16cm]{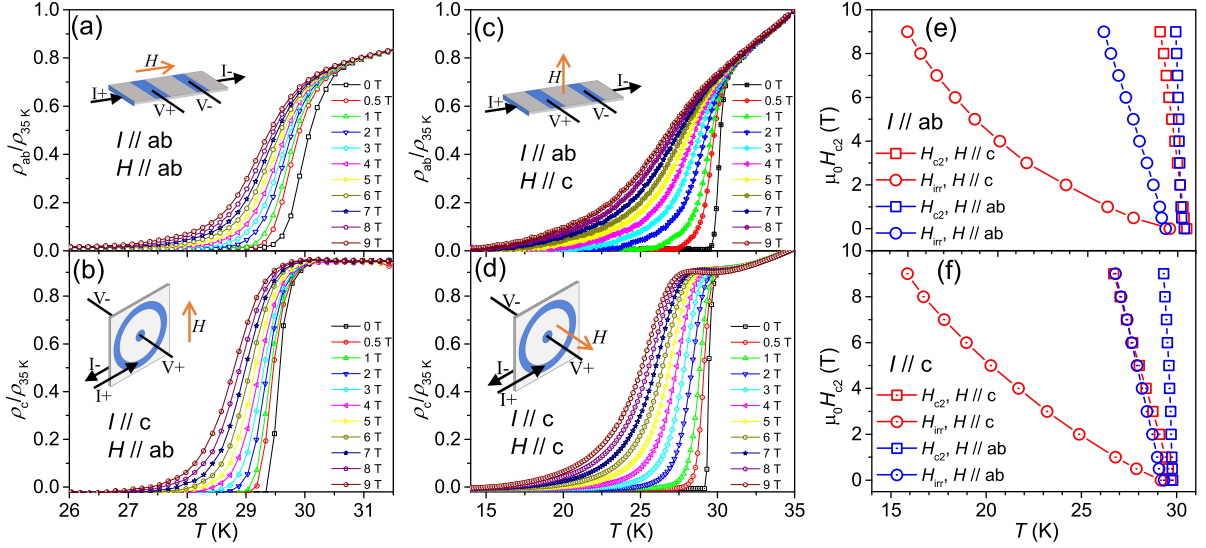}
\caption{\label{MRT}Temperature and field dependence of the in-plane and out-of-plane normalized resistivity for CsCa$_2$Fe$_4$As$_4$F$_2$ single crystals with external magnetic field parallel to the $ab$ plane (a,b) and along the $c$ axis (c,d). Panels (e) and (f) plot the upper critical field ($H_{c2}$) and the irreversible field ($H_\mathrm{irr}$) that are extracted from the $\rho_{ab}(T,H)$ and $\rho_c(T,H)$ respectively.}
\end{figure*}

Figure~\ref{RT}(a) shows the temperature dependence of the in-plane ($\rho_{ab}$) and out-of-plane ($\rho_c$) resistivity for CsCa$_2$Fe$_4$As$_4$F$_2$ single crystals at zero magnetic field. The $\rho_{ab}(T)$ behavior is quite similar to that of the polycrystalline sample~\cite{ACa2Fe4As4F2}, although the latter is about 5 times larger in magnitude. The small anomalies at high temperatures are not intrinsic, likely arising from the incomplete solidification of the silver paste on the sample. There is a gradual change in slope at around 140 K, which commonly appears in hole-doped iron-based superconductors~\cite{BaK122.Johrendt}. Below 80 K, the resistivity decreases linearly, suggesting existence of a non-Fermi-liquid state. However, one notes that the linear fit yields a negative intercept of $\rho_0=-0.0836$ m$\Omega$ cm, which seems to be unphysical because $\rho_0$ is normally referred to as a residual resistivity at zero temperature. Given the multiband feature in FeSCs, one may consider a two-fluid model with both non-Fermi liquid and Fermi liquid. According to an extended Matthiessen's rule, the total conductivity (1/$\rho_{ab}$) is contributed from the two fluids, namely,
\begin{eqnarray}
1/\rho_{ab}=1/(\rho_{1}+A_1T)+1/(\rho_{2}+A_2T^2),
\label{eq:RT}
\end{eqnarray}
where $\rho_{1}$ and $\rho_{2}$ denote the residual resistivity due to different impurity scattering within non-Fermi liquid and Fermi liquid, respectively. The data fitting in the temperature range of 32 K $<T<$ 80 K with Eq.~\ref{eq:RT} yields $\rho_{1}=$ 0.578(7) m$\Omega$ cm, $\rho_{2}=$ 0.00126(2) m$\Omega$ cm, $A_1=$ 0.0020(6) m$\Omega$ cm/K, and $A_2=$ 0.000105(3) m$\Omega$ cm/K$^2$. The result avoids any negative residual resistivity. Importantly, the fitted line extrapolated down to low temperatures [see Fig.~\ref{RT}(b)] tends to coincide with the following magnetoresistivity data [Fig.~\ref{MRT}(c)].

The $\rho_{c}(T)$ behavior is very different, both in magnitude and in the temperature dependence. The absolute resistivity at room temperature is 570 m$\Omega$ cm, 730 times larger then the $\rho_{ab}$ value at 300 K. Plotted in Fig.~\ref{RT}(d) is the resistivity anisotropy ratio, $\Gamma_{\rho}=\rho_{c}(T)/\rho_{ab}(T)$. One sees that $\Gamma_{\rho}$ increases with decreasing temperature, achieving a very large value of about 3150 at $T\rightarrow T_\mathrm{c}$. As a comparison, the $\Gamma_{\rho}$ value in BaFe$_2$As$_2$ was reported to be $\sim$100~\cite{Ba122.cxh} or even as low as $\sim$4~\cite{122.Canfield}. The giant $\Gamma_{\rho}$ is comparable to those of the Tl- and Bi-based cuprate superconductors~\cite{BiSrCaCuO,TlBaCaCuO} and electron-doped iron-selenide superconductors (Li$_{0.84}$Fe$_{0.16}$)OHFe$_{0.98}$Se~\cite{FeSe.dxl} and Li$_x$(NH$_3$)$_y$Fe$_2$Se$_2$~\cite{FeSe.LeiHC} with thick spacer layers. The result can be understood by the quasi-2D Fermi-surface sheets revealed by the first-principles calculations~\cite{12442.Hosono,12442LDA1.WangGT,12442LDA3}.

Impressively, the $\rho_{c}(T)$ curve shows a broad hump at around 140 K, reminiscence of the metal-to-nonmetal crossover in other layered metals, such as Sr$_2$RuO$_4$~\cite{Sr2RuO4,Sr2RuO4-2}, NaCo$_2$O$_4$~\cite{NaCo2O4,Na0.3CoO21.3H2O} and (Bi$_{1-x}$Pb$_x$)$_2$M$_3$Co$_2$O$_y$ (M = Ba or Sr)~\cite{Bi2M3Co2Oy,BiSrCoO}. A similar behavior was also reported in the electron-doped Li$_x$(NH$_3$)$_y$Fe$_2$Se$_2$~\cite{FeSe.LeiHC} as well as in the cuprate superconductor Tl$_2$Ba$_2$CaCu$_2$O$_8$~\cite{rho_Tl2Ba2CaCu2O8}. The metal-to-nonmetal crossover is mostly explained in terms of incoherent hopping due to $l_c\leq d_{\mathrm{inter}}$ ($l_c$ is the electronic mean free path along the $c$ axis and $d_\mathrm{inter}$ is the inter-bilayer distance). Nevertheless, the metal-to-nonmetal crossover of $\rho_{c}(T)$ in CsCa$_2$Fe$_4$As$_4$F$_2$ may also reflect the change in the electronic state because the $\rho_{ab}(T)$ exhibits a round shape at the same temperature regime.

Figure~\ref{RT}(c) magnifies the superconducting transition in $\rho_{ab}$ and $\rho_c$. One sees that the onset transition temperature ($T_c^\mathrm{onset}$) in $\rho_{ab}$ and $\rho_c$ is different. According to the criterion of 90\%$\rho_\mathrm{n}$, where $\rho_\mathrm{n}$ is the extrapolated normal-state resistivity at $T_c$, the $T_c^\mathrm{onset}$ values for $\rho_{ab}$ and $\rho_{c}$ are 30.5 K and 29.8 K, respectively. Meanwhile, the zero-resistance temperatures are basically the same ($T_c^\mathrm{zero}=29.4$ K) for $\rho_{ab}$ and $\rho_c$. The $T_c$ values of the two samples were checked to be the same (29.4$\pm$0.1 K) by dc magnetic susceptibility measurement. Also note that the measurements of $\rho_{ab}(T)$ and $\rho_c(T)$ were carried out simultaneously on the same sample puck and, at every temperature, the data were not read until holding for 30 s. Therefore, the result (higher $T_c^\mathrm{onset}$ for $\rho_{ab}$) is intrinsic (not an artifact), which suggests significant 2D superconducting fluctuations above the bulk $T_c$.

Figures~\ref{MRT}(a-d) show the superconducting transitions in $\rho_{ab}$ and $\rho_c$ under magnetic fields parallel to the $ab$ planes and the $c$ axis, respectively. In all cases, expectedly, the superconducting transitions shift to lower temperatures with increasing magnetic field. Nevertheless, details of the suppression of superconductivity are distinct for different field directions. There are pronounced resistive tails for $H \| c$ [Figs.~\ref{MRT}(c, d)]. This is mainly because, for $H \| c$, the pancake-like vortices easily flow within $ab$ planes due to the Lorentz force, giving rise to the flux-flow resistance. By contrast, the Josephson-like vortices under $H \| ab$ tend to be intrinsically pinned by the insulating spacer layers, hence the flux-flow resistance is very much reduced. Taken 90\% and 1\% of $\rho_\mathrm{n}$ as the criteria, respectively, for determining the upper critical field $H_{c2}(T)$ and the irreversible field $H_\mathrm{irr}(T)$, the resultant $H_{c2}(T)$ and $H_\mathrm{irr}(T)$ were extracted and plotted in Figs.~\ref{MRT}(e-f). The large gaps (symbols in red) between $H_{c2}^\bot(T)$ and $H_\mathrm{irr}^\bot(T)$ dictate a large area of vortex liquid phase in the $H-T$ phase diagram, akin to the scenario in many cuprate superconductors~\cite{Blatter1994}. In comparison, the gaps (symbols in blue) between $H_{c2}^\|(T)$ and $H_\mathrm{irr}^\|(T)$ are much narrower.

In general, $H_{c2}(T)$ nearby $T_c$ is dominantly limited by an orbital pair-breaking effect~\cite{review.Cao}, hence the initial slope of $H_{c2}(T)$ gives information of orbitally limited upper critical field, $H_{c2}^{\mathrm{orb}}(T)$. Note that the fluctuation effect is significant in the resistive transition of $\rho_{ab}$, as described above, we thus only use the $H_{c2}(T)$ data extracted from $\rho_c$. The initial slopes of $H_{c2}^\|(T)$ and $H_{c2}^\bot(T)$ are given to be $-$18.2 T/K and $-$2.9 T/K, respectively, by using a linear fit [note that the $\mu_0\mathrm{d}H_{c2}^\|/\mathrm{d}T)|_{T_c}$ value is very close to that of the RbGd$_2$Fe$_4$As$_4$O$_2$ polycrystals~\cite{RbGd2Fe4As4O2}]. These slopes allow us to estimate the orbitally limited upper critical fields at zero temperature with the formula, $\mu_0 H_{c2}^\mathrm{orb}(0)=-0.73T_c(\mu_0\mathrm{d}H_{c2}/\mathrm{d}T)|_{T_c}$~\cite{WHHfit} (see Table~\ref{parameters} for the result). Then the anisotropic coherence lengths can be derived using the Ginzburg-Landau (GL) relations $\mu_0 H_{c2}^{\bot}(0)=\Phi_0/[2\pi\xi_{ab}^2(0)]$ and $\mu_0 H_{c2}^{\|}(0)=\Phi_0/[2\pi\xi_{ab}(0)\xi_c(0)]$, where $\Phi_0$ is the magnetic-flux quantum ($2.07\times10^{-15}$ Wb). The $\xi_{ab}(0)$ and $\xi_c(0)$ are calculated to be 22.9 and 3.6 \AA, respectively. The resulted $\xi_c(0)$ is significantly lower than the inter-bilayer distance, $d_\mathrm{inter}=$ 8.597(4) \AA~\cite{ACa2Fe4As4F2}, which indicates that CsCa$_2$Fe$_4$As$_4$F$_2$ is actually a quasi-2D superconductor.

\begin{table}[b]
\caption{\label{parameters}Anisotropic superconducting parameters of the CsCa$_2$Fe$_4$As$_4$F$_2$. Conventional notations and definitions are used. }
\begin{ruledtabular}
\begin{tabular}{clclc}
&Parameters& &Values (unit)&\\
 \hline
&$\mu_0\frac{\mathrm{d}H_{c2}^\|}{\mathrm{d}T}|_{T_c}$ & &$-$18.2 (T/K)&\\
&$\mu_0\frac{\mathrm{d}H_{c2}^\bot}{\mathrm{d}T}|_{T_c}$ & &$-$2.9 (T/K)&\\
&$\mu_0H_{c2}^{\|,\mathrm{orb}}(0)$ & &396 (T)&\\
&$\mu_0H_{c2}^{\bot,\mathrm{orb}}(0)$ & &63 (T)&\\
&$\frac{H_{c2}^\|}{H_{c2}^\bot}|_{T_c}$ &&6.3&\\
&$\xi_{ab}(0)$ &&22.9 (\AA)&\\
&$\xi_c$(0) & &3.6 (\AA)&\\
&$\kappa_{ab}$  &&751&\\
&$\kappa_c$ &&43&\\
&$\lambda_{ab}$(0) &&986 (\AA)&\\
&$\lambda_c$(0) &&47200 (\AA)&\\
\end{tabular}
\end{ruledtabular}
\end{table}

From the initial slopes of $H_{c2}$ for the two field directions, one obtains the anisotropic ratio, $\Gamma_{H_{c2}}=H_{c2}^{\|}/H_{c2}^{\bot}=$ 6.3, at $T\rightarrow T_c$. The $\Gamma_{H_{c2}}$ value is obviously larger than those of most FeSCs~\cite{review.YHQ,Jc-Hosono_review}. In general, the anisotropy of $H_{c2}$ means the difference in the effective mass of carriers moving within and across the $ab$ planes, $m_{ab}$ and $m_c$. According to the GL theory~\cite{Blatter1994}, $\Gamma_{H_{c2}}=\sqrt{m_{c}/m_{ab}}$, then the effective-mass anisotropic ratio is $\Gamma_m=40$. The result is qualitatively consistent with the quasi-2D characteristic in the electronic structure~\cite{12442LDA3}. Nevertheless, the $\Gamma_m$ value is still much lower than the $\Gamma_{\rho}$ at $T_c$. A similar discrepancy is seen in Li$_x$(NH$_3$)$_y$Fe$_2$Se$_2$~\cite{FeSe.LeiHC}. Note that the formula $\Gamma_{H_{c2}}=\sqrt{m_{c}/m_{ab}}$ holds in a single-band scenario with coherent electronic state. Here in CsCa$_2$Fe$_4$As$_4$F$_2$, however, there are 10 bands crossing the Fermi energy~\cite{12442LDA3}. Importantly, the metal-to-nonmetal crossover in $\rho_c(T)$ suggests that only a small portion of charge carriers are in a coherent state for the $c$-direction transport. This explains an extremely large $\rho_c$ and $\Gamma_{\rho}$.


It is worth noting that the normal-state $\rho_c(T)$ under $H\| c$ shows noticeable negative magnetoresistance near $T_c$. The resistivity under $\mu_0H=$ 7 T is about 1.5\% lower than the zero-field resistivity. Furthermore, $\rho_c(T)$ shows a weak upturn under a strong field when temperature approaches $T_c$. The two phenomena are seemingly contradictory, which needs to be clarified in the future.

\subsection{Magnetic Properties}

\begin{figure}
\includegraphics[width=8cm]{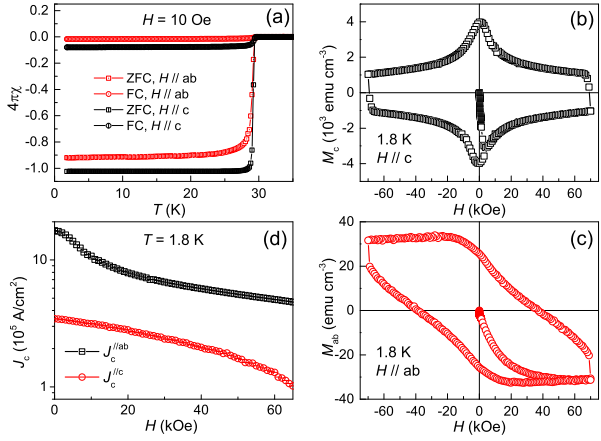}
\caption{\label{MT} (a) Temperature dependence of magnetic susceptibility of a CsCa$_2$Fe$_4$As$_4$F$_2$ single crystal under a magnetic field of $H$ = 10 Oe parallel to the $ab$ planes and the $c$ direction in field-cooling (FC) and zero-field-cooling (ZFC) modes. (b) and (c) Isothermal magnetization loops at 1.8 K for $H \| c$ and $H \| ab$, respectively. (d) Field dependence of in-plane and inter-plane superconducting critical currents at 1.8 K.}
\end{figure}

Figure~\ref{MT}(a) shows the temperature dependence of dc magnetic susceptibility of the CsCa$_2$Fe$_4$As$_4$F$_2$ single crystal, measured under a magnetic field of 10 Oe parallel to the $ab$ planes and the $c$ direction. The crystal's dimension is $A\times B\times C$ = 1.2 mm $\times$ 1.8 mm $\times$ 0.030 mm, which gives the demagnetization factors, $N_d^\|$ = 0.0367 for $H\|ab$ and $N_d^\bot$ = 0.939 for $H\|c$. With the demagnetization correction made, the volume fractions of magnetic shielding, measured in the zero-field-cooling (ZFC) mode, approximate 92\% and 102\% for $H\|ab$ and $H\|c$, respectively (exceeding 100\% is due to the uncertainties of measurements including those of the sample's dimension). The diamagnetic transition occurs at 29.4 K, in agreement with the $T_c^\mathrm{zero}$ in the resistivity measurement. The sharp diamagnetic transition as well as the nearly perfect diamagnetism demonstrates that the single crystal is of high quality. Note that the reduced diamagnetism in the field-cooling (FC) mode is due to a magnetic-flux pinning effect. The flux pinning is more obvious for $H\|ab$.

Figures~\ref{MT}(b) and (c) show the isothermal magnetization of the identical sample at 1.8 K for the two field directions, respectively. The magnetic-flux pinning effect is confirmed from the remarkable hysteresis loops from which one can estimate the critical current using Bean critical state model~\cite{Jc_BeanModel}. First, the in-plane critical current density $J_c^{\|}(H)$ can be calculated by the formula, $J_c^{\|}(H)=20\Delta M_c(H)/[A(1-A/3B)]$, where $\Delta M_c(H)$ is the magnetization difference at a certain magnetic field parallel to the $c$ direction. Second, the situation for $H\| ab$ is somewhat complicated because of the vortex motions both across and within the planes, which leads to two components for the current density, $J_c^\bot(H)$ and $J_c^{\|}(H)$. Nevertheless, $J_c^\bot$ can still be estimated by $J_c^\bot(H)\approx20\Delta M_{ab}(H)/C$~\cite{Jc_Ba2YCu3O7}, in the case of $A, B\gg(C/3)\times J_c^{\|}(H)/J_c^{\bot}(H)$ [the result is self consistent, as shown in Fig.~\ref{MT}(d)]. Under zero field, $J_c^{\|}$ and $J_c^\bot$ achieve $1.71\times10^6$ A/cm$^2$ and $3.45\times10^5$ A/cm$^2$, respectively. Even under a strong external field at $H=70$ kOe, $J_c^{\|}$ and $J_c^\bot$ remain to be as high as $4.65\times10^5$ A/cm$^2$ and $1.02\times10^5$ A/cm$^2$, respectively. These $J_c$ values are comparable to those of other iron-based superconducting thin films~\cite{Jc-Hosono_review}, suggesting potential applications at low temperatures, akin to the case of Bi-based cuprate superconductors~\cite{Blatter1994}.

\begin{figure*}
\includegraphics[width=16cm]{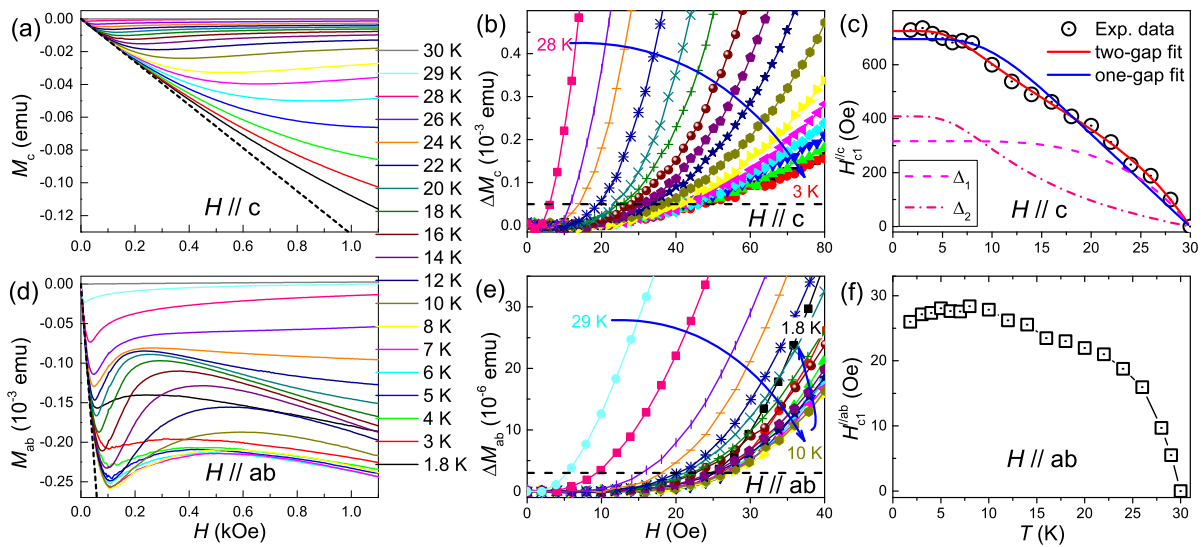}
\caption{\label{Hc1}Measurements of the anisotropic lower critical fields ($H_{c1}$) in CsCa$_2$Fe$_4$As$_4$F$_2$. (a) and (d) Field dependence of magnetization at various temperatures for $H\|c$ and $H\|ab$. The dashed lines show the linear fit in the low-field regions. Panels (b) and (e) plot the deviation of each $M(H)$ curve from the linear fit for $H\|c$ and $H\|ab$, respectively. The dashed lines are the criteria for determining $H_{c1}$. Panels (c) and (f) show the extracted and corrected (see the text for details) $H_{c1}$ for $H\|c$ and $H\|ab$, respectively. The data fittings with different models are given in (c).}
\end{figure*}

Figure~\ref{Hc1}(a) displays the $M(H)$ curves at different temperatures from 1.8 to 30 K for $H\| c$. In the low-field regions (say, $H<$ 10 Oe), the $M-H$ relation is essentially linear, as represented by the black dashed line (here we call the Meissner line). Deviations from the Meissner line dictate the field penetration into the interior of the sample, and the lower critical field ($H_{c1}$) can thus be determined. Figure~\ref{Hc1}(b) shows the deviation in magnetization $\Delta M_c$ obtained by subtracting the Meissner line from the original $M(H)$ data. We chose a criterion of $\Delta M_c=5\times10^{-5}$ emu, equivalent to 0.59 Oe after taking into account the demagnetization effect, for the determination of $H_{c1}^{\bot*}$. Those $H_{c1}^{\bot*}$ were corrected with $H_{c1}^\bot=H_{c1}^{\bot*}/(1-N_d)$ where $N_d=0.939$ for $H\|c$. The corrected $H_{c1}^\bot(T)$ data are displayed in Fig.~\ref{Hc1}(c). The zero-temperature value is  $H_{c1}^\bot(0)=$ 720 Oe, which is close to the related hole-doped superconductor Ba$_{0.6}$K$_{0.4}$Fe$_2$As$_2$\cite{Hc1_Ba0.6K0.4Fe2As2}.

$H_{c1}^\bot(T)$ data may give information on the superconducting gap~\cite{MgB2-SuperfluidDensity,Hc1_Ba0.6K0.4Fe2As2}. We tried to fit the above $H_{c1}^\bot(T)$ result considering two simple scenarios, (1) single $s$-wave-gap model and (2) two $s$-wave-gap model. $H_{c1}(T)$ relates the normalized superfluid density $\rho_s(T)$ as~\cite{MgB2-SuperfluidDensity},
\begin{eqnarray}
\frac{H_{c1}(T)}{H_{c1}(0)}=\rho_s=1+2\int_{\Delta}^{\infty}\mathrm{d}E\frac{\partial f}{\partial E}\times\frac{E}{\sqrt{E^2-\Delta(t)^2}},
\label{eq:Gapfit}
\end{eqnarray}
where $f(E)$ is Fermi function and $t$ is the reduced temperature, $T/T_c$. The temperature dependence of the BCS-like gap, $\Delta(t)$, can be approximated as $\Delta(t)=\Delta_0\tanh\{1.82[1.018(1/t-1)]^{0.51}\}$, where $\Delta(0)$ denotes the zero-temperature isotropic gap. In the two-gap scenario, the superfluid density is treated as a linear summation, $\rho_s=w\rho_s(\Delta_1)+(1-w)\rho_s(\Delta_2)$ with a weighting factor $w$.

The fitting result is shown in Fig.~\ref{Hc1}(c). Obviously, the two-gap model better meets the experimental data. The fitted parameters are: $\Delta_1(0)=2.61k_\mathrm{B}T_c=6.75$ meV, $\Delta_2(0)=0.90k_\mathrm{B}T_c=2.32$ meV, and $w=0.44$. The contribution of each superconducting gap is also presented in Fig.~\ref{Hc1}(c), revealing how the single-gap model fails to catch the experimental data. The conclusion of two superconducting gaps in CsCa$_2$Fe$_4$As$_4$F$_2$ is basically consistent with the recent muon-spin-rotation ($\mu$SR) results~\cite{KCa12442-muSR,CsCa12442-muSR}, although the latter suggest that a nodal gap should be involved. Here we cannot distinguish whether the small gap is isotropic or with nodes by the fitting. We also note that the previous measurements~\cite{KCa12442-muSR,CsCa12442-muSR} employed the polycrystalline samples with a lower $T_c$, which could lead to some differences.

The in-plane penetration depth $\lambda_{ab}$ can be estimated with the formula, $\mu_0H_{c1}^\bot=\Phi_0/(4\pi\lambda_{ab}^2)[\ln\kappa_c+0.5]$, where $\kappa_c=\lambda_{ab}/\xi_{ab}$. With $\xi_{ab}(0)=22.9$ \AA, $\lambda_{ab}(0)=$ 986 \AA~ and $\kappa_c=$ 43 are given. The $\lambda_{ab}(0)$ value is about 40\% of that derived from the $\mu$SR measurement~\cite{CsCa12442-muSR} primarily because of polycrystalline samples employed [note that $\lambda_{c}(0)$ is much larger]. One may further estimate the 2D Ginzburg number, $Gi^{\mathrm{2D}}=k_BT_c/(\sqrt2\varepsilon_0d)$, where $d$ refers to the thickness of the superconducting layers (here we take $d\approx$ 8 \AA) and  $\varepsilon_0=\Phi_0/(4\pi\lambda_{ab})^2$. The resulted $Gi^{\mathrm{2D}}$ is $0.013$, which is significantly larger than those of most FeSCs~\cite{GinzburgNumber_FeSC}, yet smaller than those of the layered Bi- and Tl-based superconductors\cite{Blatter1994,BiSrCaCuO,TlBaCaCuO}. Since $Gi^{\mathrm{2D}}$ quantifies the strength of 2D superconducting thermal fluctuations, the enhanced $T_c^\mathrm{onset}$ for $\rho_{ab}(T)$ is then naturally understood.

We also measured the isothermal magnetization curves of the CsCa$_2$Fe$_4$As$_4$F$_2$ crystal under $H \| ab$, which are shown in Fig.~\ref{Hc1}(d). Different from the $M_c(H)$ curves, the $M_{ab}(H)$ data exhibit nonmonotonic field dependence, reminiscence of the so-called fishtail effect in relation with the vortex dynamics~\cite{Fishtail,Fishtail-FeSC}. Here we only deal with the lower critical field for $H \| ab$, as what we did for $H_{c1}^{\bot}$. The criterion used for the determination of $H_{c1}^{\|*}$ is $\Delta M_{ab}=3\times10^{-6}$ emu, equivalent to 0.56 Oe. The corrected $H_{c1}^{\|}$ is close to the $H_{c1}^{\|*}$ because of the small demagnetization factor. Fig.~\ref{Hc1}(f) displays the temperature dependence of $H_{c1}^{\|}$. The $H_{c1}^{\|}(T)$ exhibits similar yet enhanced anomalies at around 25 and 8 K compared with $H_{c1}^{\bot}(T)$, suggesting a multi-gap scenario as well (the fitting on $H_{c1}^{\|}(T)$ data with Eq.~\ref{eq:Gapfit} is not meaningful because of the cylindrical Fermi-surface sheets with negligible band dispersion along the $k_z$ direction~\cite{12442LDA3}). Noticeably, $H_{c1}^{\|}(T)$ shows a maximum at around 8 K, which is very unusual for general temperature dependence of $H_{c1}$. Note that the recent $\mu$SR experiment suggests an unknown magnetic phase that appears at about 8 K~\cite{CsCa12442-muSR}, which could be the reason for the slight descending of $H_{c1}^{\|}(T)$.

The zero-temperature $H_{c1}^{\|}(0)$ extrapolated is about 25 Oe, much lower than the $H_{c1}^{\bot}(0)$ value, which also reveals the strong anisotropy in CsCa$_2$Fe$_4$As$_4$F$_2$. The inter-plane penetration depth $\lambda_c$ can be calculated with the relation, $\mu_0H_{c1}^{ab}=\Phi_0/(4\pi\lambda_{ab}\lambda_c)[\ln\kappa_{ab}+0.5]$, where $\kappa_{ab}=\sqrt{\lambda_{ab}\lambda_c}/\sqrt{\xi_{ab}\xi_c}$. The resulted $\kappa_{ab}$ and $\lambda_c(0)$, together with $\kappa_c$ and $\lambda_{ab}(0)$, are summarized in Table~\ref{parameters} for brevity. The extremely large $\lambda_c(0)$ indicates that a very low superfluid density along the $c$ direction, consistent with the quasi-2D nature of the title material.

\
\section{Concluding Remarks}

In conclusion, we have successfully grown high-quality single crystals of CsCa$_2$Fe$_4$As$_4$F$_2$, a newly discovered FeSC with separate double Fe$_2$As$_2$ layers. With the electrical and magnetic measurements we found that the material is strongly anisotropic, evidenced by the following data. (1) The normal-state resistivity anisotropy ratio is 750 at 300 K (and 3150 at 32 K); (2) In relation with (1), $\rho_c(T)$ shows a maximum of $\sim$800 m$\Omega$ cm at 140 K, and it becomes semiconducting like above 140 K, in contrast with the metallicity at the whole range in $\rho_{ab}(T)$; (3) $\rho_{ab}(T)$ shows a broadened superconducting transition with a higher $T_c^\mathrm{onset}$; (4) In the superconducting state near $T_c$, the anisotropic ratio of $H_{c2}$ is 6.3, which is higher than most FeSCs. (5) At $T\rightarrow 0$ K, the anisotropic ratio of $H_{c1}(0)$ is as high as 29; (6) The dimensionless 2D Ginzburg number achieves 0.013. All these data suggest quasi-2D characteristic in CsCa$_2$Fe$_4$As$_4$F$_2$. The result conversely implies that 2D Fe$_2$As$_2$ layers (including separate double Fe$_2$As$_2$ layers) play an important role in the emergence of superconductivity in FeSCs.

The giant anisotropy and the consequent 2D superconducting thermal fluctuations are scarcely observed in FeSCs, making the title material unique for studying the anisotropic superconductivity including the vortex dynamics. The uniqueness might be associated with the crystal structure containing separate double Fe$_2$As$_2$ layers. The special crystal structure yields the electronic structure with 6 hole-type cylindrical Fermi-surface sheets and 4 electron-type ones, all of which are very two dimensional~\cite{12442LDA3}. While two-gap model well explains the temperature dependence of the lower critical field, it is of interest for the future to further clarify which bands are responsible for superconductivity in the 12442-type system.

\begin{acknowledgments}
This work was supported by National Key Research and Development Program of China (Grant Nos. 2017YFA0303002 and 2016YFA0300202) and the Fundamental Research Funds for the Central Universities of China.
\end{acknowledgments}

\bibliographystyle{PRB}
\bibliography{12442SX}

\end{document}